\documentclass[journal]{IEEEtran}
\usepackage{cite}
\usepackage{amsmath}
\usepackage{amsfonts}
\usepackage{bm}
\usepackage{extarrows}

\usepackage{graphicx}
\usepackage[caption=false]{subfig}
\usepackage{xcolor}
\usepackage{esvect}

\newcommand{\mr}{\mathrm}

\newtheorem{proposition}{Proposition}
\newtheorem{assumption}{Assumption}

\newcommand{\cevD}[2]{\mathord{\buildrel{\lower2pt\hbox{$\scriptscriptstyle\leftarrow$}}\over{#1}}^{#2}}
\newcommand{\cevT}[3]{\mathord{\buildrel{\lower2pt\hbox{$\scriptscriptstyle\leftarrow$}}\over{#1}}^{#2}_{#3}}
\newcommand{\vecD}[2]{\mathord{\buildrel{\lower2pt\hbox{$\scriptscriptstyle\rightarrow$}}\over{#1}}^{#2}}
\newcommand{\vecT}[3]{\mathord{\buildrel{\lower2pt\hbox{$\scriptscriptstyle\rightarrow$}}\over{#1}}^{#2}_{#3}}

\begin{document}

\title{Turbo Compressed Sensing with \\Partial DFT Sensing Matrix}

\author{Junjie~Ma,
        ~Xiaojun~Yuan,~\IEEEmembership{Member,~IEEE}
        and~Li~Ping,~\IEEEmembership{Fellow,~IEEE}

\thanks{
J.~Ma and Li Ping are with the Department of Electronic Engineering, City University of Hong Kong, Hong Kong, SAR, China. (e-mail: junjiema2-c@my.cityu.edu.hk; eeliping@cityu.edu.hk.) X. Yuan is with the School of Information Science and Technology, ShanghaiTech University (email: yuanxj@shanghaitech.edu.cn).}

}

\maketitle


\begin{abstract}
In this letter, we propose a turbo compressed sensing algorithm with partial discrete Fourier transform (DFT) sensing matrices. Interestingly, the state evolution of the proposed algorithm is shown to be consistent with that derived using the replica method. Numerical results demonstrate that the proposed algorithm outperforms the well-known approximate message passing (AMP) algorithm when a partial DFT sensing matrix is involved.
\end{abstract}

\begin{IEEEkeywords}
Compressed sensing, approximate message passing (AMP), partial DFT matrix, state evolution, replica method.
\end{IEEEkeywords}

\IEEEpeerreviewmaketitle

\section{Introduction}
Partial discrete Fourier transform (DFT) sensing matrices have found many applications \cite{candes2006robust} and an efficient signal recovery algorithm is highly desirable for related compressed sensing problems. \textit{Approximate message passing} (AMP) \cite{Donoho2009,Bayati2011,Rangan2011} is an iterative algorithm for this purpose. The state evolution of AMP with independent and identically distributed (i.i.d.) Gaussian sensing matrices is shown to be consistent with that derived using the replica method \cite{Rangan2011}. This implies that AMP can potentially provide near-optimal performance when i.i.d. Gaussian sensing matrices are involved. However, the situation is different for partial DFT sensing matrices whose entries are not independently drawn. Recent results in \cite{Tulino2013,Kit2014} pointed out that, using the replica method, the optimal reconstruction performance of a system based on a partial DFT matrix is different from that based on an i.i.d. Gaussian matrix. \par
In this letter, we propose a turbo-type iterative algorithm \cite{Berrou1993} for the problem. The proposed algorithm involves two local processors. One processor handles the information related to a partial DFT sensing matrix using the linear minimum mean-square error (LMMSE) principle. The other processor handles the sparsity information. Our main contribution is a novel way to compute extrinsic messages related to the sparsity information. The state evolution of the proposed algorithm coincides with that predicted by the replica method \cite{Tulino2013,Kit2014}. This indicates the potentially excellent performance of the proposed algorithm, as confirmed by Monte Carlo simulations.
\section{Problem Description}
Consider the following linear system
\begin{equation}\label{Equ:model}
\bm{y}=\bm{F}_{\text{partial}}\bm{x}+\bm{n},
\end{equation}
where $\bm{x}\in\mathbb{C}^{N\times1}$ is a sparse signal to be estimated, $\bm{y}\in\mathbb{C}^{M\times1}$ the received signal, and $\bm{n}\sim\mathcal{CN}(\mathbf{0},\sigma^2\bm{I})$ the Gaussian noise. $\bm{F}_{\text{partial}}$ consists of $M$ randomly selected and reordered rows of the unitary DFT matrix $\bm{F}\in\mathbb{C}^{N\times N}$, where the $(m,n)$th entry of $\bm{F}$ is given by $\frac{1}{\sqrt{N}}e^{-2\pi j (m-1) (n-1)/N}$ with $j=\sqrt{-1}$. The entries of the sparse signal $\bm{x}$ is assumed to be i.i.d., with the $j$th entry of $\bm{x}$ following the Bernoulli-Gaussian distribution \cite{Rangan2011}:
\begin{equation}\label{Equ:Gauss_Markov}
x_j \sim
\begin{cases}
0 & \text{probability} = 1 - \lambda,\\
\mathcal{CN}(0,\lambda^{-1}) & \text{probability} = \lambda.
\end{cases}
\end{equation}
In \eqref{Equ:Gauss_Markov}, the variance of each $x_j$ is normalized, i.e., $\mr{E}[|x_j|^2]=1$.\par
 The partial DFT matrix in \eqref{Equ:model} can be rewritten as
\begin{equation}\label{Equ:F_partial}
\bm{F}_{\text{partial}}=\bm{SF},
\end{equation}
where $\bm{S}$ is a \textit{selection} matrix consisting of $M$ randomly selected and reordered rows of the $N\times N$ identity matrix. Define
\begin{equation}\label{Equ:u_def}
\bm{z}=\bm{Fx}.
\end{equation}
Together with \eqref{Equ:F_partial}, we rewrite the system model in \eqref{Equ:model} as
\begin{equation}\label{Equ:model_u}
\bm{y}=\bm{S}\bm{z}+\bm{n}.
\end{equation}
\subsection{Standard Turbo Algorithm}\label{Sec:conventional}
\begin{figure*}[!t]
\centering
\subfloat[A standard turbo detector.]{\includegraphics[width=.55\textwidth]{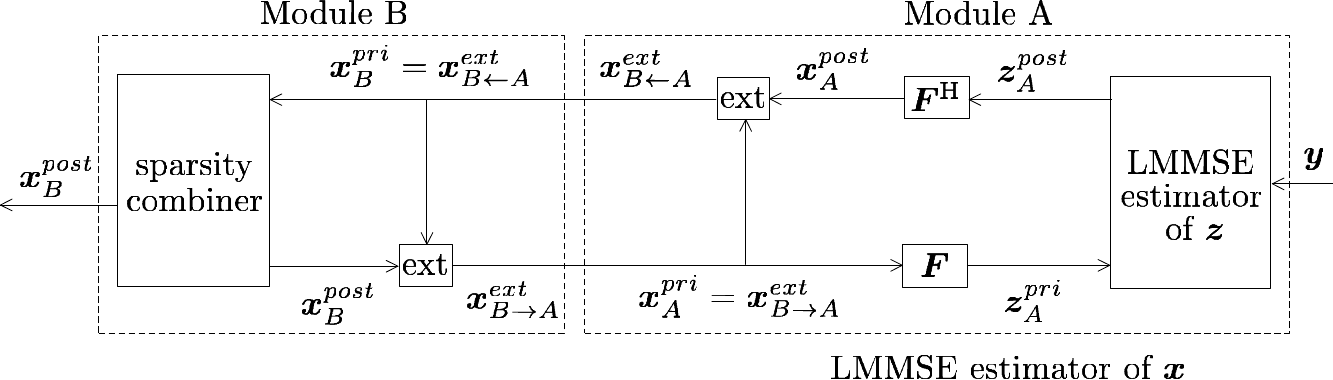}
\label{Fig:model}}
\hfil
\subfloat[The proposed turbo detector.]{\includegraphics[width=.55\textwidth]{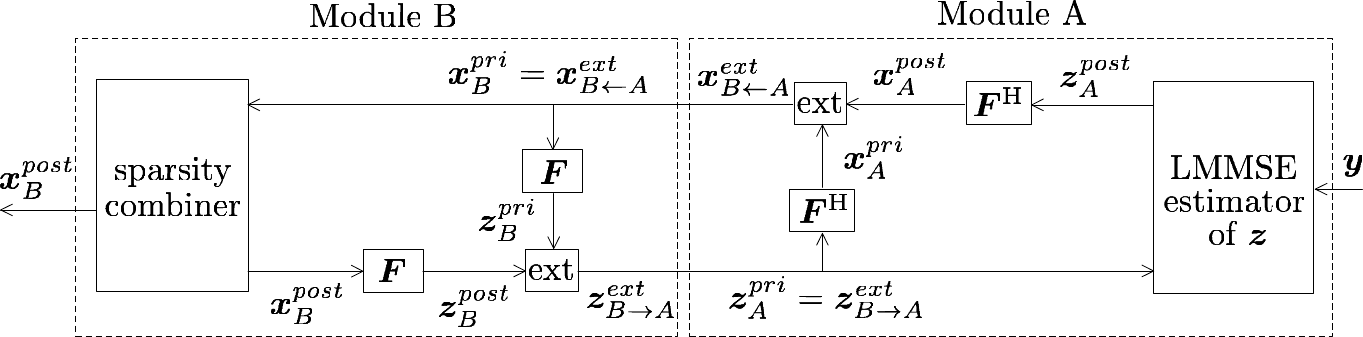}
\label{Fig:proposed}}
\caption{Block diagrams of a standard turbo algorithm and the proposed turbo algorithm. ``ext'' represents extrinsic information computation.}
\end{figure*}
The proposed algorithm is based on the turbo principle in iterative decoding \cite{Berrou1993}. Before introducing the proposed solution, we will first discuss a standard algorithm and explain its potential problem.\par
For the problem in \eqref{Equ:model}, the block diagram of a standard turbo detector is illustrated in Fig.~\ref{Fig:model}. It consists of two modules: module A is an LMMSE estimator and module B a sparsity combiner. The LMMSE estimator produces a coarse extrinsic estimate of $\bm{x}$ based on the observation $\bm{y}$. The sparsity combiner refines the estimate using the sparse distribution in \eqref{Equ:Gauss_Markov}. Here, the extrinsic output \cite{Berrou1993} of a module is fed to the other module as the \textit{a priori} input. The two modules are executed iteratively until convergence. At the end of the iteration, the final estimate of $\bm{x}$ is based on the \textit{a posteriori} output of the sparsity combiner. We next discuss the detailed operations of the algorithm in Fig.~\ref{Fig:model}.
 \\[3pt]
\textit{1) Module A:}\par
\begin{assumption}
The entries of $\bm{x}$ are i.i.d. with \textit{a priori} mean $\bm{x}_{A}^{pri}$ and variance $v_{A}^{pri}$.
\end{assumption}\par
The \textit{a priori} information about $\bm{x}$ is obtained from the feedback of the sparsity combiner,
which will be discussed later. With Assumption 1, the \textit{a priori} mean of $\bm{z}=\bm{Fx}$ is given by
\begin{equation}\label{Equ:u_pri}
\bm{z}_A^{pri}=\bm{F}\bm{x}_A^{pri}
\end{equation}
and the variance is $v_A^{pri}$. From \eqref{Equ:model_u}, the LMMSE estimator and the mean-square error (MSE) matrix of $\bm{z}$ are respectively given by \cite{Kay1993}
\begin{subequations}
\begin{align}
\bm{z}^{post}_A&=\bm{z}^{pri}_A+\frac{v_A^{pri}}{v_A^{pri}+\sigma^2}\bm{S}^{\mr{H}}\left(\bm{y}-\bm{S}\bm{z}^{pri}_A\right),\label{Equ:u_post}\\
\bm{V}_A^{post}&=v^{pri}_A\cdot\bm{I}_N-\frac{\left(v^{pri}_A\right)^2}{v^{pri}_A+\sigma^2}\bm{S}^{\mr{H}}\bm{S}.\label{Equ:u_cov}
\end{align}
\end{subequations}
From $\bm{x}=\bm{F}^{\mr{H}}\bm{z}$, the LMMSE estimator of $\bm{x}$ is
\begin{equation}\label{Equ:x_post}
\bm{x}^{post}_A=\bm{F}^{\mr{H}}\bm{z}^{post}_A.
\end{equation}
The associated MSE matrix is
\begin{equation}
\bm{F}^{\mr{H}}\bm{V}_A^{post}\bm{F}.
\end{equation}
It can be verified that the diagonals of $\bm{F}^{\mr{H}}\bm{V}_A^{post}\bm{F}$, which are the \textit{a posteriori} MSEs, are identical and given by
\begin{equation}\label{Equ:x_var_post_scalar}
v_A^{post}=v_A^{pri}-\frac{M}{N}\cdot\frac{\left(v_A^{pri}\right)^2}{v_A^{pri}+\sigma^2}.
\end{equation}
\par
Using the concise formulas in \cite{Guo2011,Xiaojun2011}, the extrinsic LMMSE estimate and the MSE of $\bm{x}$ can be computed by
\begin{subequations}\label{Equ:x_ext_all}
\begin{align}
\bm{x}^{ext}_{B \leftarrow A}&=v^{ext}_{B \leftarrow A}\left(\frac{\bm{x}_A^{post}}{v^{post}_A}-\frac{\bm{x}_A^{pri}}{v_A^{pri}}\right),\label{Equ:x_ext} \\
v_{B \leftarrow A}^{ext}&=\left(\frac{1}{v^{post}_A}-\frac{1}{v^{pri}_A}\right)^{-1}. \label{Equ:x_var_ext}
\end{align}
\end{subequations}
\\[3pt]
\textit{2) Module B:}\par
The LMMSE estimator effectively makes a Gaussian assumption on $\bm{x}$ and ignores the sparsity information of $\bm{x}$. The function of the sparsity combiner is to refine the LMMSE estimate of $\bm{x}$ by combining the sparsity information in \eqref{Equ:Gauss_Markov}.
\begin{assumption}
$\bm{x}_B^{pri}$ is modeled as an additive white Gaussian noise (AWGN) observation of $\bm{x}$, i.e.,
\begin{equation}\label{Equ:AWGN}
\bm{x}_B^{pri}=\bm{x}+\bm{w},
\end{equation}
where $\bm{w}\sim\mathcal{CN}(\mathbf{0},v_B^{pri}\bm{I})$ and is independent of $\bm{x}$. Here, $\bm{x}_B^{pri}$ and $v_B^{pri}$ are updated by the extrinsic output of module A, i.e.
\begin{equation}
\bm{x}_B^{pri} = \bm{x}_{B\leftarrow A}^{ext} \text{ and }
v_B^{pri} = v_{B\leftarrow A}^{ext}.
\end{equation}
\end{assumption}\par
Based on Assumption 2, the minimum mean-square error (MMSE) estimator of $\bm{x}$ conditioned on $\bm{x}_B^{pri}$ is a component-wise operation and given by
\begin{equation}\label{Equ:x_post_B}
{x}_{j,B}^{post}=\mr{E}\left[x_j\big|\bm{x}_B^{pri}\right]=\mr{E}\left[x_j\big|{x}_{j,B}^{pri}\right] ,\ \forall j,
\end{equation}
where ${x}_{j,B}^{post}$ and ${x}_{j,B}^{pri}$ denote the $j$th entry of $\bm{x}_{B}^{post}$ and $\bm{x}_{B}^{pri}$ respectively. $\mr{E}[\cdot]$ is with respect to the joint distribution of $\bm{x}$ and $\bm{x}_B^{pri}$ characterized by \eqref{Equ:AWGN}. The detailed operations of the above MMSE estimation can be found in, e.g., \cite{Kit2014}. The conditional variance corresponding to \eqref{Equ:x_post_B} is given by
\begin{equation}\label{Equ:x_var_post_B}
v_{j,B}^{post}=\mr{var}\left[x_j\big|\bm{x}_B^{pri}\right]=\mr{var}\left[x_j\big|{x}_{j,B}^{pri}\right],\forall j,
\end{equation}
where $\mr{var}[a|b]\equiv \mr{E}\big[|a-\mr{E}[a|b]|^2\big|b\big]$. \par
We next compute the extrinsic estimate of each $x_j$ by excluding the contribution of ${x}_{j,B}^{pri}$. Under Assumption 2, the MMSE estimation in \eqref{Equ:x_post_B} is a component-wise operation. Excluding the contribution of $x_j$, the extrinsic estimate of $x_j$ becomes
\begin{equation}\label{Equ:x_ext}
x_{j,B\to A}^{ext}=\mr{E}\big[x_j|\bm{x}_{\sim j,B}^{pri}\big]=\mr{E}\left[x_j\right]=0,\ \forall j,
\end{equation}
where $\bm{x}_{\sim j,B}^{pri}$ is obtained from $\bm{x}_{B}^{pri}$ by excluding the $j$th entry ${x}_{j,B}^{pri}$. The extrinsic estimate of module B will be treated as \textit{a priori} mean for module A in the next iteration. \par
The following observations are useful:
\begin{itemize}
\item The LMMSE operation ensures that module A in Fig.~\ref{Fig:model} is optimal (in the LMMSE sense) if the sparsity information is ignored and no iteration is involved. Note that AMP cannot make such a claim due to the distributive nature of message passing.
\item However, from \eqref{Equ:x_ext}, the extrinsic estimate of module B is zero and so iterative processing does not provide any further improvement.
\end{itemize}
In what follows, we will develop an alterative processor in Fig.~\ref{Fig:proposed} that maintains the advantage but avoid the disadvantage.
\subsection{Proposed Turbo Compressed Sensing Algorithm}
The proposed algorithm is illustrated in Fig.~\ref{Fig:proposed}. Module A computes extrinsic information of $\bm{x}$ and Module B computes the extrinsic information of $\bm{z}$. This is different from the standard approach in Fig.~\ref{Fig:model} where both modules compute extrinsic information of the same variable $\bm{x}$.
\\[3pt]
\textit{1) Module A:}\par
Module A includes the LMMSE estimator of $\bm{z}$ and two IDFTs. The operations of Module A are roughly the same as that in Fig.~\ref{Fig:model}, except that the input is $\bm{z}_{A}^{pri}$  in Fig.~\ref{Fig:proposed}.
\\[3pt]
\textit{2) Module B:}\par
As discussed in Section \ref{Sec:conventional}-2, the sparsity combiner produces no extrinsic estimate of $\bm{x}$. In the proposed algorithm, module B now computes the extrinsic estimate of $\bm{z}$ instead of $\bm{x}$.
\begin{assumption}
The \textit{a posteriori} distributions of $\bm{z}$ conditioned on $\bm{z}^{pri}_B$ are Gaussian, i.e.
\begin{equation}\label{Equ:u_post_distribution}
\mr{Pr}\left(z_j\big|\bm{z}^{pri}_B\right)=\mathcal{CN}\left(z^{post}_{j,B},v^{post}_{B}\right)\footnote{With slight abuse of notation, here (and also in \eqref{Equ:u_pri_distribution}) $\mathcal{CN}(m,v)$ denotes a Gaussian \textit{function} of $z_j$ with mean $m$ and variance $v$.},\ \forall j,
\end{equation}
where $z^{post}_{j,B}$ is the $j$th entry of the following \textit{a posteriori} mean vector
\begin{equation}\label{Equ:u_post_pri}
\bm{z}^{post}_{B}=\bm{F}\bm{x}^{post}_{B},
\end{equation}
and $v^{post}_{B}$ is the \textit{a posteriori} variance given by
\begin{equation}\label{Equ:var_post_scalar}
v^{post}_{B}=\frac{1}{N}\sum_{j=1}^N v^{post}_{j,B},
\end{equation}
\end{assumption}
where $v^{post}_{j,B}$ is the variance of $x^{post}_{j,B}$ in \eqref{Equ:x_var_post_B}.
\par
Intuitively, when $N$ is large, Assumption 3 can be justified by the mixing effect of the DFT and the central limit theorem. Eqn.~\eqref{Equ:u_post_pri} is due to $\bm{z}=\bm{Fx}$. As the entries of $\bm{x}$ are \textit{a priori} independent (from Assumption 2) and the sparsity combiner is a component-wise operation, the entries of $\bm{x}$ are also \textit{a posteriori} independent, and so \eqref{Equ:var_post_scalar} follows.\par
 From Assumption 2, the \textit{a priori} estimate $\bm{z}_B^{pri}=\bm{F}\bm{x}_B^{pri}$ is an AWGN observation of $\bm{z}$, i.e.
\begin{equation}\label{Equ:AWGN_u}
\bm{z}_B^{pri}=\bm{F}\bm{x}_B^{pri}=\bm{z}+\bm{Fw}.
\end{equation}
As $\bm{w}$ is i.i.d. Gaussian with mean zero and variance $v^{pri}_B$, $\bm{Fw}$ has the same distribution and we have
\begin{equation}\label{Equ:u_pri_distribution}
\mr{Pr}\left(z_{j,B}^{pri}\big|z_j\right)=\mathcal{CN}\left(z_{j,B}^{pri},v_{B}^{pri}\right),\ \forall j.
\end{equation}\par
From \eqref{Equ:AWGN_u}, $z_{j,B}^{pri}$ and $\bm{z}_{\sim j,B}^{pri}$ are conditionally independent given ${z}_j$. It can then be verified that
\begin{equation}\label{Equ:u_decomposition}
\mr{Pr}\left(z_j|\bm{z}_{B}^{pri}\right)\propto\mr{Pr}\left(z_j|\bm{z}_{\sim j,B}^{pri}\right)\cdot \mr{Pr}\left({z}_{j,B}^{pri}|z_j\right),\ \forall j
\end{equation}
where $\propto$ denotes equality up to a constant scaling factor independent of $z_j$. Based on \eqref{Equ:u_post_distribution}, \eqref{Equ:u_pri_distribution} and \eqref{Equ:u_decomposition}, the extrinsic distribution $\mr{Pr}(z_j|\bm{z}_{\sim j,B}^{pri})$ is Gaussian \cite{Loeliger2007,Guo2011} and given by
\begin{equation}\label{Equ:u_ext_distribution}
\mr{Pr}\left(z_j|\bm{z}_{\sim j,B}^{pri}\right)=\mathcal{CN}\left({z}_{j,B\to A}^{ext},v_{B\to A}^{ext}\right),\ \forall j
\end{equation}
where
\begin{subequations}\label{Equ:u_ext_all}
\begin{align}
z_{j,B\to A}^{ext}&=v_{B\to A}^{ext}\left(\frac{z_{j,B}^{post}}{v_{B}^{post}}-\frac{z_{j,B}^{pri}}{v_{B}^{pri}}\right), \label{Equ:u_ext} \\
v_{B\to A}^{ext}&=\left(\frac{1}{v_{B}^{post}}-\frac{1}{v_{B}^{pri}}\right)^{-1}. \label{Equ:u_ext_var}
\end{align}
\end{subequations}
The extrinsic mean/variance in \eqref{Equ:u_ext_all} will be treated as \textit{a priori} mean/variance for module A in the next iteration. Note that in the standard turbo detector in Section \ref{Sec:conventional}, module B produces no extrinsic output, as shown in \eqref{Equ:x_ext}. This is the main difference between the proposed algorithm and the standard detector.\\[3pt]
\textit{4) Overall Algorithm}\par
In the first iteration, $\bm{z}_{A}^{pri}=\mathbf{0}$ and  $v_{A}^{pri}=1$. The operations of module A and module B are executed iteratively until convergence. \par
The DFT/IDFT operations in Fig.~\ref{Fig:proposed} can be efficiently implemented using the fast Fourier transform (FFT). Also, the order of the ``ext'' operations in Fig.~\ref{Fig:proposed} (see \eqref{Equ:x_ext_all} and \eqref{Equ:u_ext_all}) and DFT/IDFT can be changed, and then one pair of DFT/IDFT can be saved. This is straightforward and we omit the details.
\section{State Evolution}\label{Sec:evolution}
Following \cite{Donoho2009,Bayati2011,Rangan2011}, we analyze the large-system performance of the proposed scheme by using state evolution.
\subsection{State Evolution}\label{Sec:evolution}
We characterize the performance of the iterative algorithm by a recursion of two states, $v_{A}^{pri}$ and $v_{B}^{pri}$. In the following, for notational brevity, we define
\begin{equation}\label{Equ:def}
\eta\equiv\frac{1}{v_B^{pri}}\text{ and }\ v\equiv v_A^{pri}.
\end{equation}
We define the following MMSE of the sparse signal estimation given an AWGN observation (with SNR $\eta$)
\begin{equation}\label{Equ:MMSE_AWGN}
mmse(\eta)\equiv \mr{E}\left[|x-\mr{E}[x|x+\xi]|^2\right],
\end{equation}
 where $x$ is a sparse signal modeled as \eqref{Equ:Gauss_Markov} and $\xi\sim\mathcal{CN}(0,\eta^{-1})$. From \eqref{Equ:x_var_post_B} and based on Assumption 2,
\begin{equation}\label{Equ:limit}
v_{B}^{post}=\frac{1}{N}\sum_{j=1}^N\mr{var}\left[x_j|x_{j,B}^{post}\right]\to mmse(\eta).
\end{equation}
Under Assumptions 1-3, we have the following proposition.
\begin{proposition}\label{Pro:1}
The state evolution of the proposed turbo compressed sensing algorithm is characterized by
 \begin{subequations}\label{Equ:fixed}
\begin{align}
\eta_{t+1}&=\frac{1}{\frac{N}{M}\cdot(v_{t}+\sigma^2)-v_{t}},\label{Equ:fixed_a}\\
\frac{1}{v_{t+1}}&=\frac{1}{mmse(\eta_{t+1})} - \eta_{t+1},\label{Equ:fixed_b}
\end{align}
\end{subequations}
\end{proposition}
where the subscript $t$ and $t+1$ indicate the iteration indices. The state evolution in \eqref{Equ:fixed} is derived by combining \eqref{Equ:x_var_post_scalar}, \eqref{Equ:x_var_ext}, \eqref{Equ:x_var_post_B}, \eqref{Equ:var_post_scalar}, \eqref{Equ:u_ext_var}, \eqref{Equ:def}-\eqref{Equ:limit}, together with some straightforward manipulations.
\subsection{Fixed Point of State Evolution}
 Denote by $\eta_{\star}$ the convergence value of $\eta$. Combining \eqref{Equ:fixed_a} and \eqref{Equ:fixed_b} and eliminating $v$, $\eta_{\star}$ can be characterized by the following fixed point equation:
\begin{equation}\label{Equ:fixed2}
\frac{N}{M}\cdot\sigma^2\cdot mmse(\eta_{\star})\cdot\eta_{\star}^2-\frac{N}{M}\cdot\left(mmse(\eta_{\star})+\sigma^2\right)\cdot\eta_{\star}+1=0.
\end{equation}
One solution of \eqref{Equ:fixed2} is given by
\begin{equation}\label{Equ:solution}
\eta_{\star}=\frac{\left(mmse+\sigma^2\right)-\sqrt{\left(mmse+\sigma^2\right)^2-4\sigma^2\cdot mmse\cdot\frac{M}{N}}}{2\cdot\sigma^2\cdot mmse},
\end{equation}
where $mmse$ represents $mmse(\eta_{\star})$. Note that \eqref{Equ:fixed2} has two solutions, but it can be shown that the other solution is not a valid convergence point. \par
It can be verified that \eqref{Equ:solution} is consistent with that in \cite[(17) and (37)]{Tulino2013} derived using the replica method. It can also be shown that \eqref{Equ:fixed} is equivalent to \cite[(17)-(18)]{Kit2014}. We omit the details here due to space limitation.
\section{Numerical Examples}\label{Sec:numerical}
\begin{figure}[htbp]
\centering
  \includegraphics[width=.4\textwidth]{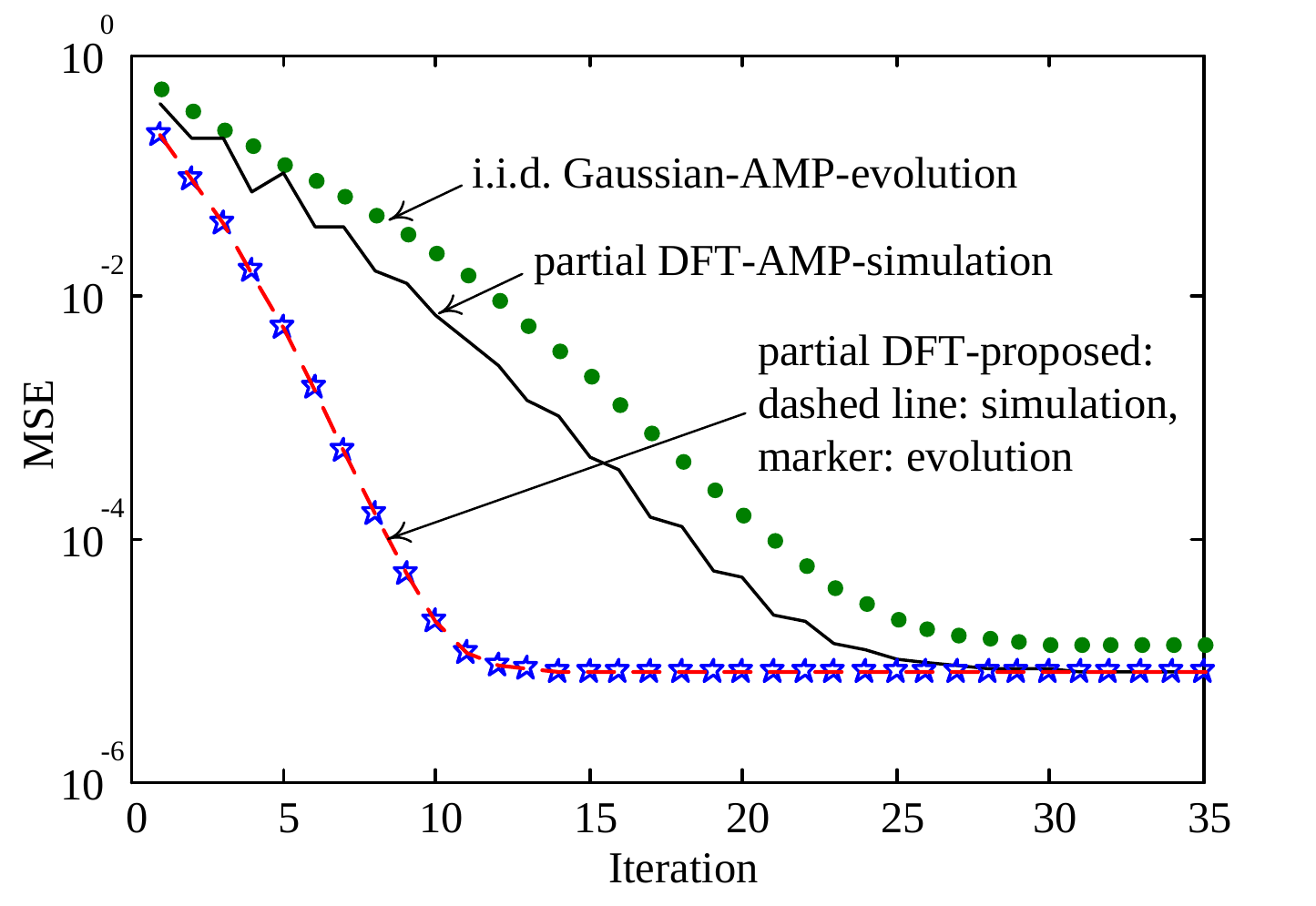}
  \caption{Comparisons of the proposed algorithm and AMP. $N = 8192$, $M = 5734\,(\approx0.7 N)$, $ \lambda = 0.4$, and SNR = 50 dB.}\label{Fig:numerical}
\end{figure}
\begin{figure}[htbp]
\centering
  \includegraphics[width=.4\textwidth]{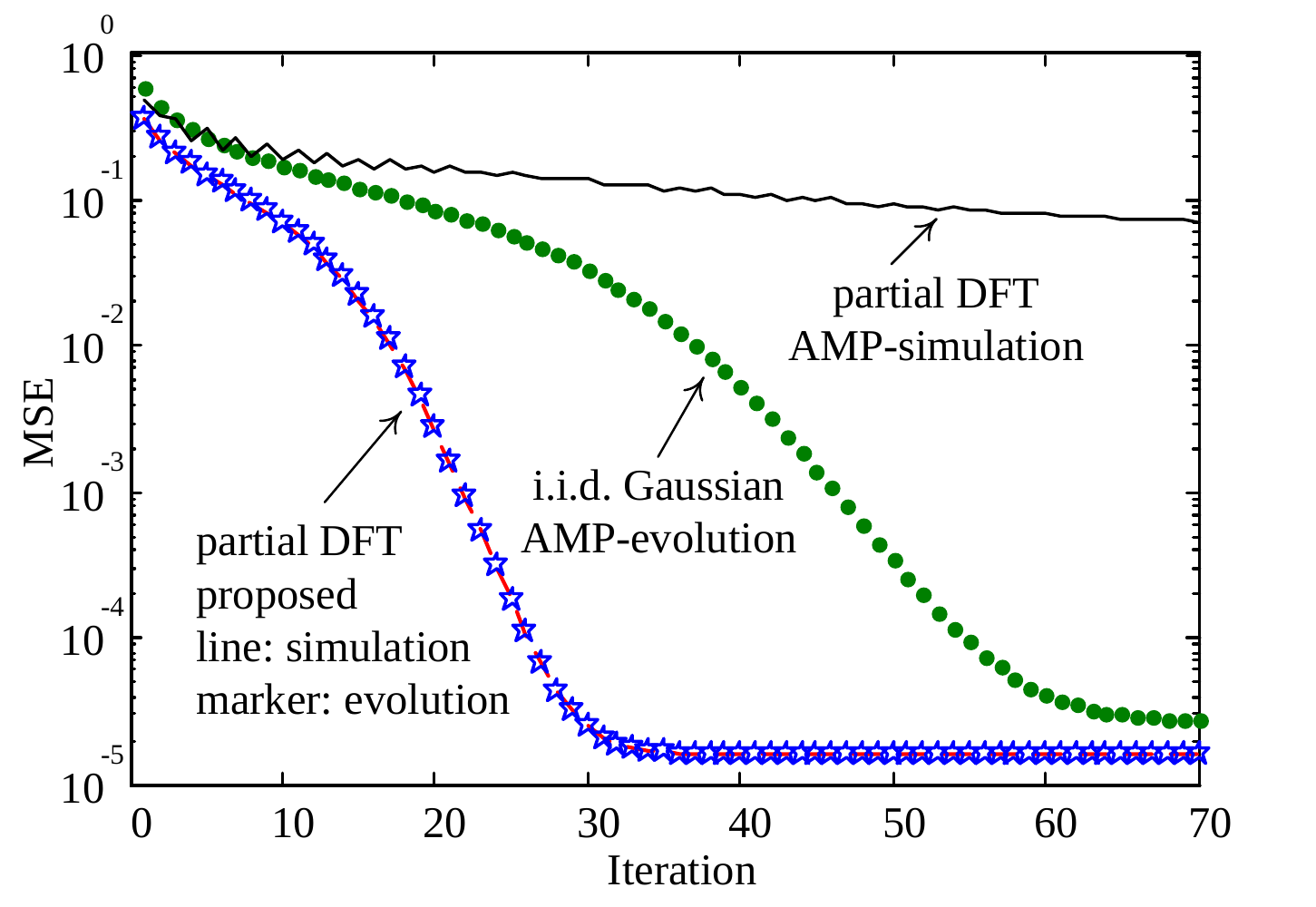}
  \caption{Comparisons of the proposed algorithm and AMP. $N = 32768$, $M = 18022\,(\approx0.55 N)$, $ \lambda = 0.4$, and SNR = 50 dB.}\label{Fig:numerical2}
\end{figure}
In Fig.~\ref{Fig:numerical}, we compare the MSE performance of the proposed algorithm with partial DFT matrices and AMP with i.i.d. Gaussian matrices. For a fair comparison, the variance of each entry in the i.i.d. Gaussian matrix is normalized to $1/N$. Here, the implementation of AMP is based on \cite{Rangan2011}. In simulation, MSE is obtained by averaging over 2000 realizations. We see that the proposed algorithm converges faster than AMP and also achieves lower convergence MSE. Moreover, the state evolution analysis agrees well with simulation. We can also directly apply AMP to the case with a partial DFT sensing matrix. From Fig.~\ref{Fig:numerical}, we see that AMP with partial DFT matrices outperforms the case with i.i.d. Gaussian matrices. This performance difference also indicates that the state evolution of AMP is not accurate when applied to partial DFT matrices. This is reasonable because the state evolution of AMP is developed for i.i.d. Gaussian matrices.\par
In Fig.~\ref{Fig:numerical2}, we reduce the measurement ratio $M/N$. ($N$ is set to sufficiently large so that the simulation performance agrees well with state evolution.) We see that AMP for partial DFT performs much worse than the proposed algorithm in this setup.
\section{Conclusion and Discussions}
The state evolution in Section \ref{Sec:evolution} is developed based on three assumptions. Numerical results in Section \ref{Sec:numerical} demonstrate that the state evolution developed based on these assumptions is accurate. It is an interesting future research topic to establish more rigorous justifications for the state evolution. The analysis in \cite{Bayati2011} for AMP with i.i.d. Gaussian sensing matrices may shed light on this problem.

\ifCLASSOPTIONcaptionsoff
  \newpage
\fi
\bibliographystyle{IEEEtran}	
\bibliography{IEEEabrv,CompressedSensing}		

\begin{thebibliography}{10}
\providecommand{\url}[1]{#1}
\csname url@samestyle\endcsname
\providecommand{\newblock}{\relax}
\providecommand{\bibinfo}[2]{#2}
\providecommand{\BIBentrySTDinterwordspacing}{\spaceskip=0pt\relax}
\providecommand{\BIBentryALTinterwordstretchfactor}{4}
\providecommand{\BIBentryALTinterwordspacing}{\spaceskip=\fontdimen2\font plus
\BIBentryALTinterwordstretchfactor\fontdimen3\font minus
  \fontdimen4\font\relax}
\providecommand{\BIBforeignlanguage}[2]{{%
\expandafter\ifx\csname l@#1\endcsname\relax
\typeout{** WARNING: IEEEtran.bst: No hyphenation pattern has been}%
\typeout{** loaded for the language `#1'. Using the pattern for}%
\typeout{** the default language instead.}%
\else
\language=\csname l@#1\endcsname
\fi
#2}}
\providecommand{\BIBdecl}{\relax}
\BIBdecl

\bibitem{candes2006robust}
E.~J. Cand{\`e}s, J.~Romberg, and T.~Tao, ``Robust uncertainty principles:
  Exact signal reconstruction from highly incomplete frequency information,''
  \emph{{IEEE} Trans. Inf. Theory}, vol.~52, no.~2, pp. 489--509, 2006.

\bibitem{Donoho2009}
D.~L. Donoho, A.~Maleki, and A.~Montanari, ``Message-passing algorithms for
  compressed sensing,'' \emph{Proceedings of the National Academy of Sciences},
  vol. 106, no.~45, pp. 18\,914--18\,919, 2009.

\bibitem{Bayati2011}
M.~Bayati and A.~Montanari, ``The dynamics of message passing on dense graphs,
  with applications to compressed sensing,'' \emph{{IEEE} Trans. Inf. Theory},
  vol.~57, no.~2, pp. 764--785, Feb 2011.

\bibitem{Rangan2011}
\BIBentryALTinterwordspacing
S.~Rangan. Generalized approximate message passing for estimation with random
  linear mixing. Preprint, 2010. [Online]. Available:
  \url{http://arxiv.org/abs/1010.5141}
\BIBentrySTDinterwordspacing

\bibitem{Tulino2013}
A.~Tulino, G.~Caire, S.~Verdu, and S.~Shamai, ``Support recovery with sparsely
  sampled free random matrices,'' \emph{{IEEE} Trans. Inf. Theory}, vol.~59,
  no.~7, pp. 4243--4271, July 2013.

\bibitem{Kit2014}
\BIBentryALTinterwordspacing
C.~K. Wen and K.~K. Wong. Analysis of compressed sensing with spatially-coupled
  orthogonal matrices. Preprint, 2014. [Online]. Available:
  \url{http://arxiv.org/abs/1402.3215}
\BIBentrySTDinterwordspacing

\bibitem{Berrou1993}
C.~Berrou and A.~Glavieux, ``Near optimum error correcting coding and decoding:
  turbo-codes,'' \emph{{IEEE} Trans. Commun.}, vol.~44, no.~10, pp. 1261--1271,
  Oct 1996.

\bibitem{Kay1993}
S.~M. Kay, \emph{Fundamentals of Statistical Signal Processing: Estimation
  Theory}.\hskip 1em plus 0.5em minus 0.4em\relax NJ: Prentice-Hall PTR, 1993.

\bibitem{Guo2011}
Q.~Guo and D.~Huang, ``A concise representation for the soft-in soft-out lmmse
  detector,'' \emph{{IEEE} Commun. Lett.}, vol.~15, no.~5, pp. 566--568, May
  2011.

\bibitem{Xiaojun2011}
\BIBentryALTinterwordspacing
X.~Yuan, L.~Ping, C.~Xu, and A.~Kavcic. Achievable rates of mimo systems with
  linear precoding and iterative lmmse detection. Preprint, 2011. [Online].
  Available: \url{http://arxiv.org/ftp/arxiv/papers/1106/1106.0178.pdf}
\BIBentrySTDinterwordspacing

\bibitem{Loeliger2007}
H.-A. Loeliger, J.~Dauwels, J.~Hu, S.~Korl, L.~Ping, and F.~Kschischang, ``The
  factor graph approach to model-based signal processing,'' \emph{Proceedings
  of the IEEE}, vol.~95, no.~6, pp. 1295--1322, June 2007.

\end{thebibliography}

\end{document}